\documentclass[aps,pra,superscriptaddress,reprint,floatfix,longbibliography,showpacs]{revtex4-1}
\usepackage[english]{babel}
\usepackage{amsmath}
\usepackage{amssymb}
\usepackage{graphicx}
\usepackage{upgreek}
\usepackage{epstopdf}
\usepackage{color}
\usepackage{dcolumn}
\usepackage{multirow}
\usepackage[note-name=, use-sort-key = false]{notes2bib}

\usepackage[colorlinks,bookmarks=false,citecolor=darkblue,linkcolor=red,urlcolor=blue]{hyperref}
\definecolor{darkred}{rgb}{0.7,0.0,0.0}
\definecolor{darkblue}{rgb}{0,0.02,0.45}

\begin{document}
\title{Optical conductivity of the metallic pyrochlore iridate Pr$_2$Ir$_2$O$_7$: Influence of spin-orbit coupling and electronic correlations on the electronic structure}

\author{Harish Kumar}
\email{harish.kumar@physik.uni-augsburg.de}
\affiliation{Experimental Physics II, Institute of Physics, University of Augsburg, 86159 Augsburg, Germany}

\author{M. K\"opf}
\affiliation{Experimental Physics II, Institute of Physics, University of Augsburg, 86159 Augsburg, Germany}

\author{P. Telang}
\affiliation{Experimentalphysik VI, Center for Electronic Correlations and Magnetism, University of Augsburg, 86159 Augsburg, Germany}

\author{N. Bura}
\affiliation{Experimental Physics II, Institute of Physics, University of Augsburg, 86159 Augsburg, Germany}

\author{A. Jesche}
\affiliation{Experimentalphysik VI, Center for Electronic Correlations and Magnetism, University of Augsburg, 86159 Augsburg, Germany}

\author{P. Gegenwart}
\affiliation{Experimentalphysik VI, Center for Electronic Correlations and Magnetism, University of Augsburg, 86159 Augsburg, Germany}

\author{C. A. Kuntscher}
\email{christine.kuntscher@physik.uni-augsburg.de}
\affiliation{Experimental Physics II, Institute of Physics, University of Augsburg, 86159 Augsburg, Germany}

\begin{abstract}
The synergy of strong spin-orbit coupling and electron-electron interactions gives rise to unconventional topological states, such as topological Mott insulator, Weyl semimetal, and quantum spin liquid.
In this study, we have grown single crystals of the pyrochlore iridate Pr$_2$Ir$_2$O$_7$ and explored its magnetic, lattice dynamical, and electronic properties. While Raman spectroscopy data reveal six phonon modes confirming the cubic \textit{Fd$\bar{3}$m} crystal symmetry, dc magnetic susceptibility data show no anomalies and hence indicate the absence of magnetic phase transitions down to 2~K.
Both temperature-dependent electric transport and optical conductivity data reveal the metallic character of Pr$_2$Ir$_2$O$_7$.
The optical conductivity spectrum contains a mid-infrared absorption band, which becomes more pronounced with decreasing temperature due to spectral weight transfer from high to low energies. The presence of the mid-infrared band hints at the importance of correlation physics. The optical response furthermore suggests that Pr$_2$Ir$_2$O$_7$ is close to the Weyl semimetal phase.

\end{abstract}

\maketitle

\section{Introduction}
Pyrochlore iridates $A$$_2$Ir$_2$O$_7$, with $A$ = Y, Bi and rare earths $R$, have garnered significant interest due to their distinctive properties, such as frustration, the presence of iridium, and the potential for novel topological phases \cite{Pesin.2010,William.2014}. The substantial spin-orbit coupling (SOC) arising from iridium's heavy nature, coupled with the extended 5$d$ orbitals minimizing electronic correlation ($U$) effects, contributes to the intriguing nature of these materials. In iridates, SOC, $U$, and crystal electric field effect (CEF) exhibit comparable strength, and balancing these energies can lead to the emergence of distinct topological phases \cite{Pesin.2010,William.2014,Wan.2011}. The CEF induces the splitting of 5$d$ orbitals into $t_{2g}$ and $e_{g}$ states, and the robust SOC divides the $t_{2g}$ level into a $J_{eff}$ = 1/2 doublet and a $J_{eff}$ = 3/2 quartet. In the pyrochlore iridates the Ir has a 4+ valency with a 5$d^5$ configuration, resulting in a spin-orbital moment $J_{eff}$ = 1/2 scenario. The small $U$ leads to the $J_{eff}$ = 1/2 splitting into a lower and an upper Hubbard band, introducing interesting physics, related to the spin-orbit Mott insulator state.
Recent theoretical investigations proposed intriguing topological states for pyrochlore iridates \cite{Pesin.2010,William.2014,Wan.2011,Moon.2013}. Initially, Wan and others introduced a Weyl semimetal (WSM) phase for the magnetic pyrochlores, featuring an all-in/all-out antiferromagnetic (AFM) order that preserves structural symmetry while breaking time-reversal symmetry \cite{Wan.2011}. Depending on the strength of $U$, either a Mott insulator with an all-in/all-out magnetic structure or a WSM phase with 24 Weyl nodes at the Fermi surface was predicted \cite{Wan.2011}.

$R$-pyrochlore iridates containing 5$d$ electrons (Ir$^{4+}$) generally demonstrate an insulating AFM state, where this behavior is typically suppressed with an increase in the size of the $R$-ion \cite{Matsuhira.2011,Nakatsuji.2006,Harish.2016,harish.2021,Taira.2001}. Accordingly, Pr$_2$Ir$_2$O$_7$ is not insulating, but displays a paramagnetic (PM) metallic nature \cite{Nakatsuji.2006,harish.2021,conf.2018}. This is further confirmed by optical studies, where systems containing Y, Dy, Eu, and Sm demonstrate a distinct optical gap ($\Delta_{g}$ $>$ 0.2 eV) that diminishes with larger $A$-cations, as evidenced by Nd and Pr materials \cite{Ueda.2016, Ueda.2012}.
Namely, Nd$_2$Ir$_2$O$_7$ contains unconventional free charge, evidenced by the quadratic temperature dependence of the Drude spectral weight, which is consistent with the characteristics expected for massless Dirac electrons. However, it is notable that the entropy counterpart does not exhibit a $T^3$ contribution in the specific heat \cite{Wang.2020}.
Moreover, the substituted material Nd$_2$(Ir$_{1-x}$Rh$_x$)$_2$O$_7$ demonstrates a transition from a narrow-gap Mott insulator to a WSM and ultimately to a correlated metal as $x$ increases \cite{Ueda.2012}. In addition, the optical response of Eu$_2$Ir$_2$O$_7$ exhibits a linear-in-frequency behavior at low temperatures, suggesting the characteristic signature of the WSM phase \cite{Sushkov.2015}.
In other pyrochlore iridates, infrared studies of Bi$_2$Ir$_2$O$_7$ and Pb$_2$Ir$_2$O$_7$ show a strongly metallic ground state and a sizable mid-infrared absorption band at around 0.2 eV and 0.4 eV, respectively \cite{Lee.2013,Hirata.2013}.
The presence of an AFM state in most pyrochlore iridates breaks time-reversal symmetry while preserving crystal symmetries, however, Pr$_2$Ir$_2$O$_7$ is an exceptional case among the $R$ pyrochlores \cite{William.2014,Nakatsuji.2006,harish.2021}.

\begin{figure}[t]
	\includegraphics[width=1\linewidth]{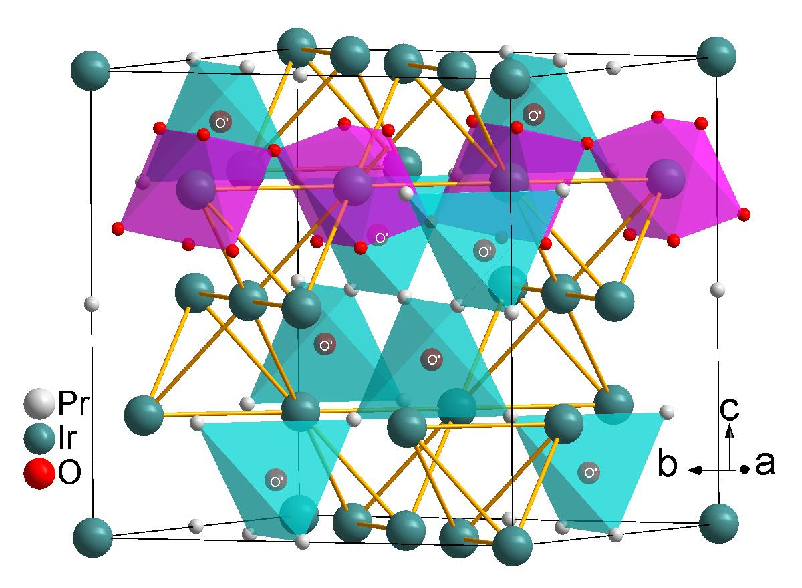}
	\caption{The cubic pyrochlore structure shows corner-shared tetrahedral arrangements of Pr$_4$O$^{'}$ (in cyan) and Ir$_4$$\diamondsuit$ (in orange edges), where $\diamondsuit$ represents empty center-site (8a-site) \cite{Harish.2016}. Additionally, the structure includes IrO$_6$ octahedra (in light magenta).}
	\label{fig.Cubic}
\end{figure}

For Pr$_2$Ir$_2$O$_7$, theoretical predictions propose a quadratic band touching at the zone center \cite{Goswami.2017}, which has been confirmed by angle-resolved photoemission spectroscopy experiments \cite{Kondo.2015}.
The presence of a magnetic and sizable Pr$^{3+}$ (4$f^2$) ion adds to its distinctive nature. Pr$_2$Ir$_2$O$_7$ consists of two magnetically active sublattices, Pr and Ir, forming corner-shared tetrahedra (see Fig.\ \ref{fig.Cubic}) and introducing magnetic frustration. Interestingly, an AFM Ruderman–-Kittel–-Kasuya–-Yosida (RKKY) interaction with an energy scale of $T_N$ $\sim$ 20 K between Pr-4$f$ moments, mediated through Ir-5$d$ delocalized electrons, has been observed \cite{Nakatsuji.2006}. The AFM interaction experiences a suppression due to the screening of 4$f$ moments through the Kondo effect, resulting in a reduction of the Weiss temperature to $\mathrm{\theta_{CW}}$ = 1.7 K. Notably, below 1.7 K, the susceptibility of Pr$_2$Ir$_2$O$_7$ exhibits an $\ln T$ dependence \cite{Nakatsuji.2006} and thermodynamic measurements indicate a quantum critical scaling, interpreted as signature the metallic spin liquid phase \cite{Tokiwa.2014}.

Additionally, an unusual anomalous Hall effect (AHE) has been identified below 1 K in Pr$_2$Ir$_2$O$_7$ which is elucidated by the spin-chirality effect in Ir-5$d$ electrons, arising from the noncoplanar spin structure of Pr spins \cite{Machida.2007}. The noteworthy observation of AHE, even in the absence of uniform magnetization at zero field, provides compelling evidence for the existence of the long-sought chiral spin liquid state in Pr$_2$Ir$_2$O$_7$ \cite{Machida.2009}. More intriguingly, from a theoretical perspective, this material is proposed to be situated in proximity to an interaction-driven AFM quantum critical point (QCP) \cite{Savary.2014} compatible with the experimental signatures \cite{Tokiwa.2014}.
This QCP has been considered as a pivotal transition point between an AFM WSM and a nodal non-Fermi liquid phase \cite{Savary.2014}. Furthermore, Ueda \textit{et. al} identified distinct signatures of topological transitions within the WSM states for Pr$_2$Ir$_2$O$_7$ under the influence of a magnetic field, prominently attributed to the $f-d$ coupling \cite{KUeda.2022}.

In this work, we present the synthesis of Pr$_2$Ir$_2$O$_7$ single crystals and their comprehensive characterization. In Sections A, B, and C, we present the magnetic and charge transport properties, as well as Raman spectroscopy results, which align with previous findings. However, our main focus lies in investigating the optical properties of Pr$_2$Ir$_2$O$_7$. The optical conductivity spectrum contains a Drude contribution revealing the metallic character of Pr$_2$Ir$_2$O$_7$ and high-energy excitations due to transitions between $J_{eff}$ = 3/2 and $J_{eff}$ = 1/2 bands.
Furthermore, an absorption band in the midinfrared frequency range whose spectral weight is growing with decreasing temperature suggests the importance of electronic correlations.
The results are compared with recent findings for closely related metallic pyrochlore iridates and discussed in terms of correlated semimetals.

\section{EXPERIMENTAL DETAILS}
Single crystals of Pr$_2$Ir$_2$O$_7$ were grown using polycrystalline material synthesized via the KF flux method within a platinum crucible. The polycrystalline precursor was prepared by combining Pr$_{6}$O$_{11}$ and Ir powders in a stoichiometric ratio, each with a phase purity exceeding 99.99\% (M/s Sigma-Aldrich). The Pr$_{6}$O$_{11}$ underwent a pre-heat treatment at 800$^{\circ}$C for $\sim$ 8 h to eliminate residual atmospheric moisture. The thoroughly ground mixture of Pr$_6$O$_{11}$, and Ir, further mixed with $\sim$ 2\% potassium fluoride (KF) by mixture's mass, was further ground to obtain a homogeneous composition. This well-mixed powder was then pelletized and sintered in air at a temperature of 800$^{\circ}$C for 48 hours, involving an intermediate grinding step to enhance the material's homogeneity and crystalline structure. The phase purity of the material was checked with powder x-ray diffraction (XRD) using a Rigaku diffractometer with CuK$_{\alpha}$ radiation.

\begin{figure}[t]
	\includegraphics[width=1\linewidth]{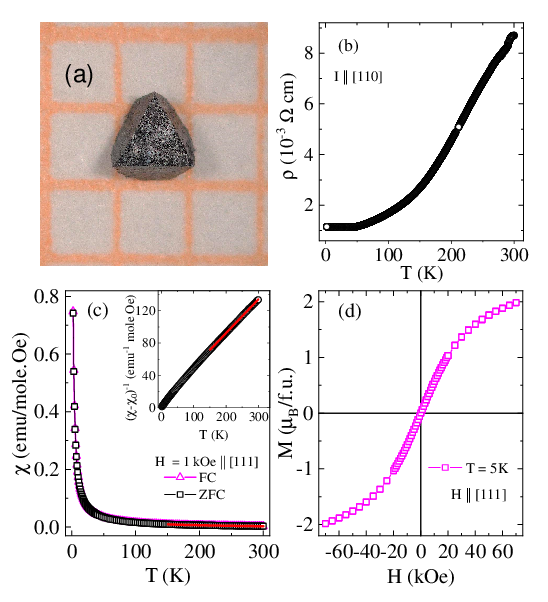}
	\caption{(a) Image of a Pr$_2$Ir$_2$O$_7$ single crystal. (b) dc electrical resistivity $\rho(T)$ data of Pr$_2$Ir$_2$O$_7$ as a function of temperature.  (c) Temperature-dependent magnetic susceptibility data of Pr$_2$Ir$_2$O$_7$, where the solid line in the high-temperature regime is fit to Eq. 1. Inset of (c) shows the inverse magnetic susceptibility $(\chi - \chi_0)^{-1} (T)$ data with a straight-line fitting in the high-temperature regime. (d)  Magnetic field-dependent magnetization of Pr$_2$Ir$_2$O$_7$ measured at 5 K.}
	\label{fig.characterization1}
\end{figure}

To synthesize the single crystals, the obtained polycrystalline Pr$_2$Ir$_2$O$_7$ material is blended with KF flux and the excess of IrO$_2$, and subjected to calcination in a platinum crucible, and comprehensive details of crystal growth are provided in Refs. \cite{MILLICAN.2007,Kumar.2024}. Within the Pt-crucible, a collection of octahedron-shaped crystals emerges at the bottom. Subsequently, the crystals are carefully extracted from the crucible using distilled hot water, revealing a characteristic size of approximately $\sim$ 1 mm, as illustrated in Fig.\ \ref{fig.characterization1}(a). The chemical composition of the Pr$_2$Ir$_2$O$_7$ crystals was verified using energy dispersive analysis of x-ray (EDX) on a ZEISS Crossbeam 550/550L scanning electron microscope, equipped with an Oxford detector.
The obtained atomic percentages of Pr and Ir in Pr$_2$Ir$_2$O$_7$ are 19.8\% and 19.4\%, respectively. Previous studies have shown that stoichiometry significantly impacts the electronic properties of R$_2$Ir$_2$O$_7$, as confirmed by the ratio of Pr and Ir \cite{Sleight.2018}. In Pr$_2$Ir$_2$O$_7$, the ratio of Pr and Ir is found to be 1.02, which closely aligns with the nominal concentration.
Raman spectroscopy measurements were performed on Pr$_2$Ir$_2$O$_7$ single crystals utilizing a confocal micro-Raman setup featuring a Jobin–Yvon T64000 spectrograph and a 488 nm argon ion laser as excitation source was employed.

Temperature and field-dependent magnetization data have been collected using a SQUID magnetometer (MPMS). Electrical transport properties have been investigated in the temperature range of 2 to 300 K using a four-probe technique with PPMS from Quantum Design.
Temperature-dependent infrared spectroscopy (IR) measurements were conducted with unpolarized electromagnetic radiation in the frequency range of 210 to 18000 cm$^{-1}$ (26 meV to 2.23 eV) within a temperature range of 300 to 5 K. This was done using a CryoVac Konti cryostat connected to a Bruker Vertex 80v FTIR spectrometer, coupled with a Bruker Hyperion infrared microscope. For reflectivity measurements, the (111) crystal surface was polished with diamond paper.
To obtain absolute reflectivity, a silver layer was deposited onto half of the polished crystal surface, serving as a reference \cite{Kopf.2022,Kumar2022}. Subsequently, the sample was fixed to a sample holder within the cryostat using low-temperature GE-varnish, ensuring its alignment perpendicular to the incident beam. In the far-infrared range, the frequency resolution was set to 1 cm$^{-1}$, while a resolution of 4 cm$^{-1}$ was selected for the higher-frequency range. The obtained reflectivity spectrum was extrapolated in the low- and high-energy regimes using Drude-Lorentz fitting and x-ray optic volumetric data, respectively \cite{Tanner.2015}. It's important to note that a power-law ($\omega^{n}$) interpolation with an integer $n$ = 3 was applied to bridge the region between measured reflectivity and calculated high-energy extrapolated data. From the resulting reflectivity spectrum the real part of the optical conductivity $\sigma_{1}$($\omega$) was calculated through Kramers-Kronig relations. Further, the Drude-Lorentz model was adeptly applied for fitting the optical spectra, facilitated by the RefFIT program \cite{Kuzmenko.2005}.

\section{Results}

\subsection{Magnetic properties}
Figure\ \ref{fig.characterization1}(c) depicts the temperature-dependent magnetization $M(T)$ data for Pr$_2$Ir$_2$O$_7$, recorded from 2 K to 300 K under an applied fields H = 1 kOe along the [111] direction following zero-field cooled (ZFC) and field cooled (FC) protocol. The magnetic moment increases as temperature decreases, with a notable increase below around 30 K, consistent with prior studies \cite{Nakatsuji.2006,harish.2021,conf.2018}. The absence of magnetic irreversibility between ZFC and FC magnetization suggests a PM behavior for this material. A representative fit of the $\chi(T)$ data with modified Curie Weiss behavior (Eq.\ \ref{equ-curie}) in the temperature range of 150 - 300 K is presented in Fig.\ \ref{fig.characterization1}(c) as a solid line \cite{harish.2017,harish.2021,Nakatsuji.2006}.
\begin{eqnarray} \label{equ-curie}
\chi(T) = \chi_0 + \frac{C}{(T - \theta_{CW})} \quad,
\end{eqnarray}
where the variables $\chi_0$, $C$, and $\mathrm{\theta_{CW}}$ represent the temperature-independent magnetic susceptibility, Curie constant, and Curie Weiss temperature, respectively.
The reasonably good fit with Eq.\ \ref{equ-curie} indicates that the magnetic state in the PM state obeys modified Curie-Weiss behavior. The fitting yields a Curie temperature $\mathrm{\theta_{CW}}$ = - 22.4 K, consistent with prior studies \cite{Nakatsuji.2006,harish.2021}. The corrected inverse susceptibility $(\chi-\chi_0)^{-1}$ versus T exhibits a linear behavior in the high-temperature regime (Inset of Figure\ \ref{fig.characterization1}c).

Remarkably, the emergence of a finite $\mathrm{\theta_{CW}}$ for Pr$_2$Ir$_2$O$_7$, in the absence of a Neel temperature ($T_N$), is important to note. This finite $|$$\mathrm{\theta_{CW}}$$|$ occurs due to Ir-5$d$ electrons mediated AFM-type RKKY interactions among Pr-4$f$ atoms \cite{Nakatsuji.2006}. This observation finds further support in recent experiments involving the substitution of magnetic Pr at the nonmagnetic Y-site in (Y$_{1-x}$Pr$_x$)$_2$Ir$_2$O$_7$, where the magnitude of $\mathrm{\theta_{CW}}$ rapidly weakens with increasing Pr content \cite{harish.2021}. It is believed that this reduction in $|$$\mathrm{\theta_{CW}}$$|$ is attributed to the $f-d$ interaction between localized Pr-4$f$ and itinerant Ir-5$d$ electrons. Using the value of the Curie constant $C$ = 2.4, we have calculated the effective magnetic moment $\mathrm{\mu_{eff}}$ according to $\mathrm{\mu_{eff}}$ = $\sqrt{3k_{B}C/(N_{A}\mu^{2}_{B})}$ = 2.83$\sqrt{C}$. The so-obtained value of $\mathrm{\mu_{eff}}$ = 4.38 $\mathrm{\mu_{B}/f.u.}$ is quite close to the reported results  \cite{Nakatsuji.2006,harish.2021}.

The magnetic field dependence data $M(H)$ measured at 5 K in the field range of $\pm$ 70 kOe applied along [111] is depicted in Fig.\ \ref{fig.characterization1}(d). As evident in the figure, the magnetic moment exhibits a non-linear increase with increasing field, with no signs of saturation observed up to 70 kOe. The moment recorded at 5 K under the highest measuring field (70 kOe) is approximately 1.98 $\mathrm{\mu_{B}/f.u.}$, consistent with previously reported values \cite{MacHida.2005,harish.2021}.

\begin{figure}[t]
	\includegraphics[width=1\linewidth]{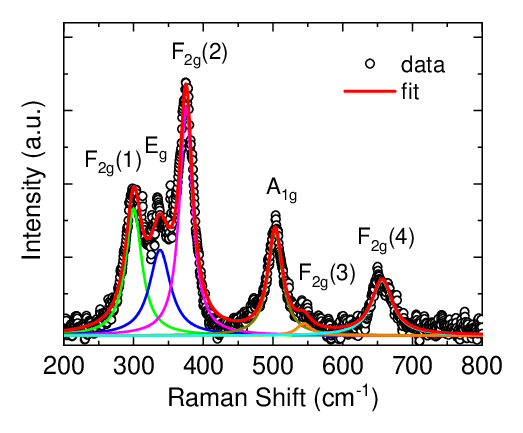}
	\caption{Raman spectrum for Pr$_2$Ir$_2$O$_7$ at room temperature together with the total fit and the individual phonon mode contributions.}
	\label{fig.Raman}
\end{figure}

\subsection{Electric transport}

To comprehend the electronic conduction characteristics of Pr$_2$Ir$_2$O$_7$, we performed dc resistivity $\rho(T)$ measurements as a function of temperature in a four-probe geometry. As evident in Fig.\ \ref{fig.characterization1}(b),
$\rho(T)$ monotonically decreases with decreasing temperature down to $\sim$ 45 K evidencing metallic behavior. Between $\sim$ 45 K and 25 K $\rho(T)$ changes only slightly, followed by a subtle increase below 25 K down to 2 K.
The decrease in resistivity with lowering the temperature with a minimum at around 30~K is
consistent with pervious studies \cite{Nakatsuji.2006,harish.2021,conf.2018,Takatsu.2014} and it follows a logarithm (ln$T$) dependence (not shown), indicative of the Kondo effect in Pr$_2$Ir$_2$O$_7$ arising due to $f-d$ interaction between localized Pr-4$f$ magnetic ions and itinerant Ir-5$d$ electrons \cite{Nakatsuji.2006,harish.2021}.


\subsection{Raman Spectroscopy}
To confirm the cubic \textit{Fd$\bar{3}$m} symmetry of Pr$_2$Ir$_2$O$_7$, Raman spectroscopy has been employed. Room-temperature Raman spectroscopy data of Pr$_2$Ir$_2$O$_7$ are depicted in Fig.\ \ref{fig.Raman}. According to the cubic phase with \textit{Fd$\bar{3}$m} symmetry, six Raman active modes are expected based on the factor group analysis \cite{Keimer.2019,Kumar.2023}:
\begin{eqnarray}
        \mathrm{\Gamma_{optic} =  [A_{1g} + E_{g} + 4 F_{2g}]_{R} +  [7 T_{1u}]_{IR}} \quad ,
\end{eqnarray}
where R and IR indicate the Raman and infrared active modes, respectively.
In Fig.\ \ref{fig.Raman}, six distinctive Raman modes of Pr$_2$Ir$_2$O$_7$ are observed and tabulated in Table\ \ref{tab:table 1}. The spectrum characteristics and the position of the modes confirm the cubic \textit{Fd$\bar{3}$m} symmetry and align seamlessly with prior theoretical and experimental studies, as indicated in Refs.\ \cite{Keimer.2019,Kumar.2023,Rosalin.2023}.

\begin{table}[b]
\caption{\label{tab:table 1} Raman modes at 300~K and T$_{1u}$ infrared-active modes (labelled $L_1$...$L_4$) observed at 300 K and 5 K. }
\begin{ruledtabular}
\begin{tabular}{ccccccc}
\underline{Raman Modes (cm$^{-1}$)}
\\
T(K) & F$_{2g}^1$ & E$_{g}$ & F$_{2g}^2$ & A$_{1g}$ & F$_{2g}^3$  & F$_{2g}^4$ \\
\hline
300 K & 300.2 & 338.0 & 375.4 & 503.2 & 544.3 & 656.4\\
\hline
\underline{Infrared Modes (cm$^{-1}$)}
\\
T(K) &  $L_1$ & $L_2$ & $L_3$ & $L_4$ &  & \\
\hline
300 K  & 338.1 & 408.4 & 463.5 & 575.7 & &\\
5 K  & 350.4 & 407.7 & 451.3 & 579.8 & &\\
\end{tabular}
\end{ruledtabular}
\end{table}

\begin{figure}[t]
	\includegraphics[width=1\linewidth]{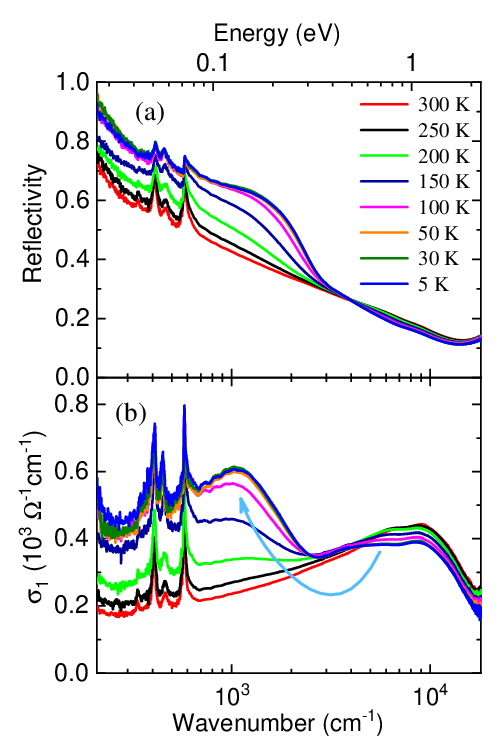}
	\caption{(a) Reflectivity and (b) optical conductivity $\sigma_1$ spectra of Pr$_2$Ir$_2$O$_7$ as a function of temperature. The curved arrow in (b) highlights the spectral weight transfer from high to low frequencies with cooling.}
	\label{fig.reflectivity1}
\end{figure}

\subsection{Infrared Spectroscopy}
To comprehend the evolution of the electronic band structure, we conducted reflectivity measurements on the (111) crystal surface covering a broad frequency range. Fig.\ \ref{fig.reflectivity1}(a) shows the temperature-dependent reflectivity spectrum for various temperatures. From the room temperature reflectivity spectrum $R(\omega)$ we identified four infrared (IR)-active phonon modes in the measured far-infrared region and list them in Table\ \ref{tab:table 1}. The $L_1$, $L_2$, $L_3$, and $L_4$  IR-modes are associated with the motion of the Ir–-O–-Ir bond. Specifically, $L_1$ corresponds to bending (out-of-plane) mode, while $L_2$ and $L_3$ correspond to bending (in-plane) mode, and $L_4$ is related to the stretching mode \cite{Son.2019}. The positions of the observed IR modes are consistent with both theoretical and experimental values \cite{Keimer.2019}. $R(\omega)$ intriguingly rises with decreasing temperature in the mid and far-infrared regime, displaying a heightened level in the far-IR region at 5 K and approaches unity at the lowest frequencies, indicating the metallic nature for Pr$_2$Ir$_2$O$_7$. The temperature-dependent $R(\omega)$ spectra of Pr$_2$Ir$_2$O$_7$ exhibit a notable hump or plasma edge-type characteristic around the mid-IR region ($\sim$ 0.15 eV), followed by a gradual reduction in reflectivity as the frequency extends up to 2 eV.


The corresponding optical conductivity spectra $\sigma_{1}(\omega)$, which derived from the reflectivity spectra through Kramers-Kronig (KK) analysis, are depicted in Fig.\ \ref{fig.reflectivity1}(b).
The total spectral weight is conserved in the frequency range 0 - 18.000 cm$^{-1}$ for all measured temperatures.
At room temperature, the $\sigma_{1}(\omega)$ spectra display a Drude term at low frequency due to intraband transitions, suggesting the metallic nature of Pr$_2$Ir$_2$O$_7$. Additionally, four IR modes are observed in the far-IR region, and a mid-IR absorption peak, along with four interband features is observed in the high-frequency region(to be discussed later).
Analysis of temperature-dependent $\sigma_{1}(\omega$) spectra reveals several key observations: 1) the optical conductivity undergoes marked changes with temperature. A clear rise in conductivity is evident in the far-IR region with decreasing temperature, consistent with the metallic nature of Pr$_2$Ir$_2$O$_7$. Notably, the magnitude of optical conductivity at the lowest measured frequencies is consistent with the measured dc conductivity using the four-probe method (discussed later). 2) the most noticeable change in the optical conductivity at 300 K is the low energy part (mid-IR region) of the spectrum which composed a broad and prominent $M$-band at $\sim$0.12~eV [see Fig.\ \ref{fig.reflectivity1}(b)] and it becomes more prominent with lowering the temperature. 3) There is a spectral weight ($SW$) transfer from the high energy side to the low energy side (mid-IR) with lowering the temperature, as indicated by the curved arrow in Fig.\ \ref{fig.reflectivity1}(b), passing through an isosbestic point at around $\sim$ 4000 cm$^{-1}$. A similar type of spectral weight transformation has been seen in A$_2$Ir$_2$O$_7$ when transitioning from insulating (Y, Dy, Eu) to metallic (Nd, Pr) phases \cite{Ueda.2016}.

\begin{figure}[t]
	\includegraphics[width=1\linewidth]{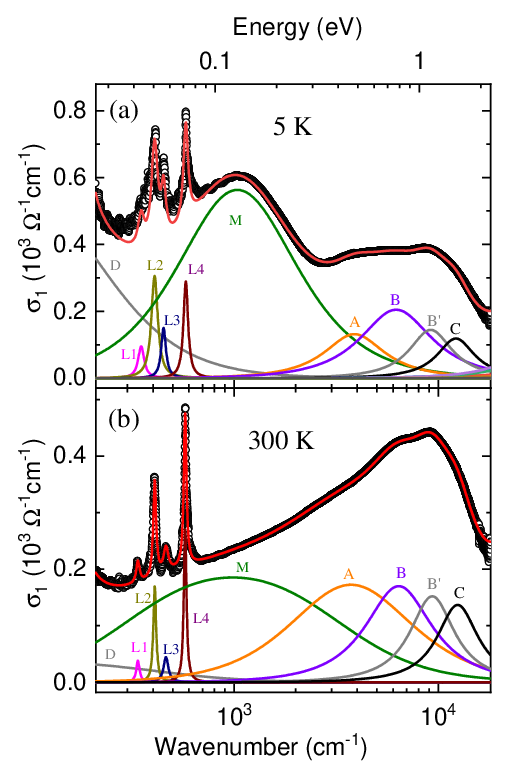}
	\caption{Optical conductivity $\sigma_1$ spectrum at (a) 5 K and (b) 300 K together with the Drude-Lorentz fit and the fit contributions.}
	\label{fig.Fitting1}
\end{figure}

\begin{figure}[t]
	\includegraphics[width=0.9\linewidth]{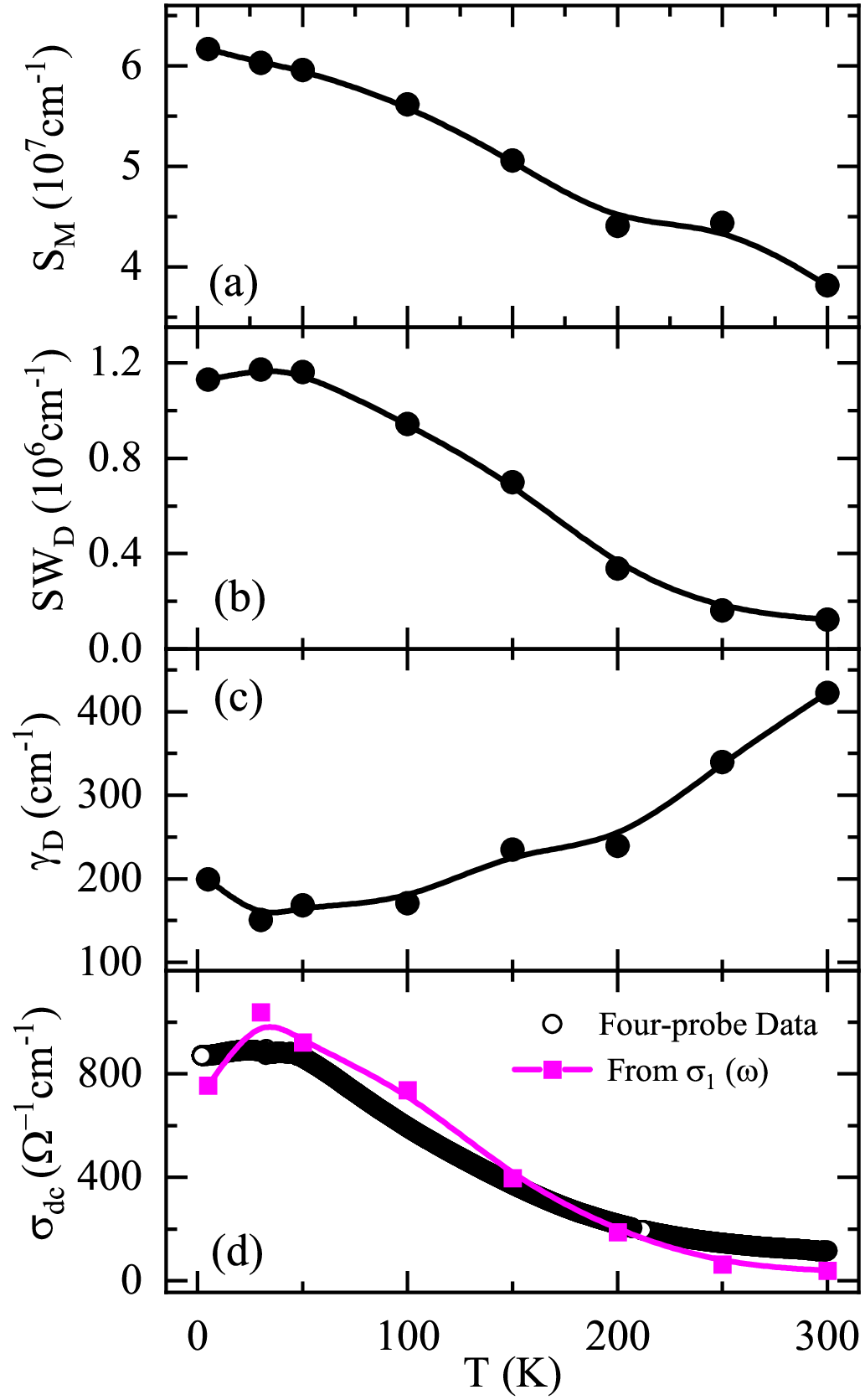}
	\caption{(a) Mode strength $S_M$ of the $M$-band, (b) spectral weight $SW_{D}$ of the Drude term, (c) scattering rate $\gamma_{D}$ of the Drude term, and (d) dc conductivity $\sigma_{dc}$ obtained from four-probe method and extracted from $\sigma_{1}(\omega$) data, respectively, as a function of temperature.}
	\label{fig.parameters1}
\end{figure}

For quantitative analysis, we employed the Drude-Lorentz model to reproduce the $\sigma_{1}(\omega$) spectrum, where the Drude component describes the free charge carrier excitations, and the Lorentz terms is associated with phonon excitations and interband transitions. The general formula for the Drude-Lorentz model is expressed as follows \cite{Dressel.2002}:
\begin{eqnarray}
\sigma_1(\omega) = \frac{\omega_{p}^2}{4\pi}\frac{\tau_{D}}{1 + \omega^2 \tau_{D}^2} + \sum_{j} \frac{S_j}{4\pi}\frac{\omega^2/\tau_{j}}{(\omega_{j}^2 - \omega^2)^2 + \omega^2/\tau_{j}^2} \quad
\end{eqnarray}
where the first term represents the Drude contribution and the latter terms the Lorentz-type excitations. Within the Drude term, the $\omega_{p}$ and 1/$\tau_{D}$ denote the plasma frequency and relaxation rate of individual conduction bands. The scattering rate $\gamma_{D}$ of Drude peak can be derived from the relaxation rate using $\gamma_{D}$ = 1/(2$\pi c\tau_D$), where $c$ represents the speed of light. Meanwhile, the plasma frequency can defined as $\omega^2_P = 4 \pi ne ^2 / m^* $, where $n$ is the carrier concentration and $ m^* $ is the effective mass of the charge carriers. In the second term, $\omega_{j}$, 1/$\tau_{j}$, and S$_j$ are indicative of the resonance frequency, broadening of the oscillator (width = 1/$2 \pi c \tau_{j}$), and mode strength for each Lorentz oscillator, respectively.  An essential aspect of conductivity is its spectral weight $SW$ = $\int{\sigma_{1}(\omega)}d{\omega}$, defined as the integral of the optical conductivity over specific frequency intervals.
Integrating $\sigma_{1}(\omega)$ over finite frequency ranges, particularly where distinct features such as interband excitations are observed, quantifies the number of electrons involved in particular features like interband excitation processes.
For the Drude term, the $SW_{D}$ is related to the plasma frequency according to $\omega^{2}_{p}$/8 \cite{Dressel.2002} and will be discussed later in Fig.\ \ref{fig.parameters1}.

\begin{figure}[t]
	\includegraphics[width=0.8\linewidth]{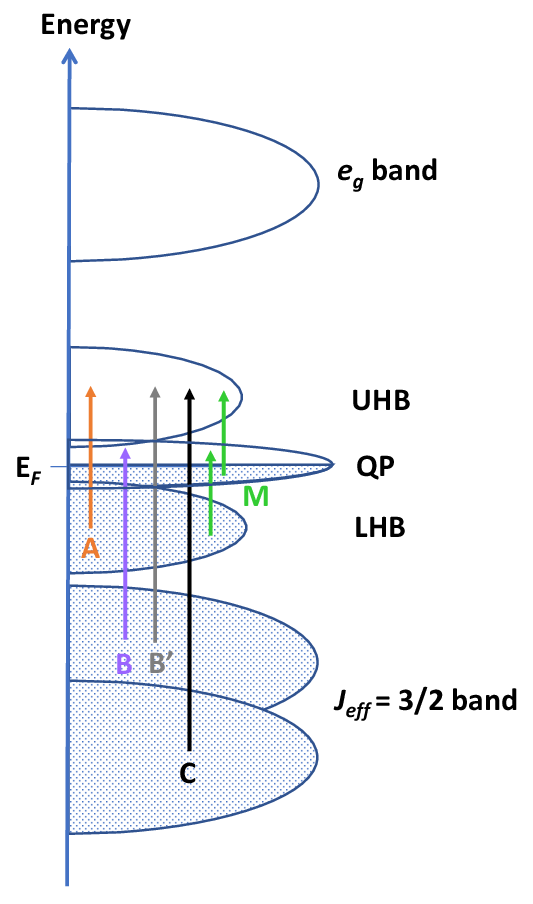}
	\caption{Suggested energy band scheme of Pr$_2$Ir$_2$O$_7$ based on its optical conductivity. The labelled arrows $B$, $B^{'}$, and $C$ correspond to transitions from $J_{eff,3/2}$ to $J_{eff,1/2}$, while $A$ is attributed to excitations between the Hubbard bands. Additionally, the $M$ denotes transitions either from the lower Hubbard band to the QP peak or from the QP peak to the upper Hubbard band.}
	\label{fig.Scheme}
\end{figure}

Considering the metallic nature of Pr$_2$Ir$_2$O$_7$, we incorporated one Drude term ($D$) for excitations of itinerant carriers, along with five Lorentz oscillators labelled $M$, $A$, $B$, $B^{'}$, and $C$ corresponding to electronic transitions.
Hall transport investigations revealed that Pr$_2$Ir$_2$O$_7$ predominantly exhibits electron-type charge carriers, characterized by a carrier concentration of $n$ = 4.13$\times$10$^{21}$ cm$^{-3}$ \cite{Machida.2007}. Consequently, a single Drude term is employed to characterize the optical conductivity.
Figs.\ \ref{fig.Fitting1}(a) and (b) illustrate the optical conductivity ($\sigma_{1}$) of Pr$_2$Ir$_2$O$_7$ in the entire measured frequency range, showing representative Lorentz fitting and individual Lorentz contributions at 5 K and 300 K, respectively. The observed IR modes at 5 K and 300 K for Pr$_2$Ir$_2$O$_7$ are shown in Fig.\ \ref{fig.Fitting1} by using $L_{1}$, $L_{2}$, $L_{3}$ and $L_{4}$ notations and are tabulated in Table \ref{tab:table 1}.

In the high-energy $\sigma_{1}(\omega$) spectrum, four distinct Lorentz contributions are identified at $\sim$0.46 eV ($A$), $\sim$0.79 eV ($B$), $\sim$1.15 eV ($B^{'}$), and $\sim$1.54 eV ($C$).
Consistent with other iridates \cite{Kim.2008,Kumar2022,Hermann.2017}, the last three excitations ($B$, $B^{’}$ and $C$) are associated with transitions between $J_{eff}$ = 3/2 and $J_{eff}$ = 1/2 states, exhibiting potential additional splittings due to octahedral distortion \cite{Kumar2022}, as illustrated in the energy band scheme depicted in Fig.\ \ref{fig.Scheme}.
The observed peak $A$ around $\sim$0.46 eV closely aligns with reported values for iridates \cite{Kim.2008,Moon.2008,Kumar2022}. This spectral feature predominantly arises from transitions within the $J_{eff}$ = 1/2 band split by electron correlations. It's noteworthy that with decreasing temperature, these excitations experience a slight shift (approximately 0.02 eV) towards the low-energy side, except the A peak, which shifts to higher energy by a similar amount. We highlight the pivotal role of strong SOC, which prevents the merging of these interband features into a single transition from $t_{2g}$ to $e_{g}$ bands. Consequently, we deduce that SOC plays a pivotal role in shaping the electronic structure of Pr$_2$Ir$_2$O$_7$.

Besides these excitations, the $M$-band and the Drude contribution ($D$) exhibit interesting changes with temperature. Figs.\ \ref{fig.parameters1}(a)--(d) show the temperature-dependent mode strength of $M$-band ($S_M$), spectral weight of Drude ($SW_{D}$), scattering rate of Drude ($\gamma_{D}$), and the dc conductivity data ($\sigma_{dc}$) obtained from four-probe method and optical $\sigma_{1}(\omega$) data, respectively.
The strength $S_M$ of the $M$-band and the spectral weight of the Drude term $SW_{D}$ both increase with lowering the temperature [see Figs.\ \ref{fig.parameters1}(a) and (b), respectively]. The scattering rate $\gamma_{D}$ of the itinerant charge carriers decrease with decreasing temperature [Fig.\ \ref{fig.parameters1}(c)] and amounts to 200 cm$^{-1}$ at 5~K. This value yields a scattering time $\tau$ of about 2.6$\times$10$^{-14}$ s, derived from the formula $\gamma_{D}$ = 1/(2$\pi$c$\tau$).
Employing the obtained $\tau$ value, the carrier mobility $\mu$ is calculated using the expression $\mu$ = e$\tau$/$m^*$, where $m^{*}$ is the effective carrier mass. A recent photoemission study of Pr$_2$Ir$_2$O$_7$ has revealed an effective carrier mass ($m^{*}$) of 6.3 times the mass of a free electron ($m_0$)\cite{Kondo.2015}. Subsequently, the resulting mobility is estimated at around 7.4 cm$^{2}$/Vs, a value in close agreement with the reported electrical transport mobilities in single crystals ($\sim$4.19 cm$^{2}$/Vs)\cite{Machida.2007,Nakatsuji.2006} and thin films ($\sim$28 cm$^{2}$/Vs)\cite{Ohtsuki.2019}.

It is important to note that both the mode strength of $M$-band and spectral weight of the Drude term ($D$) increase with lowering the temperature.
The M-band in the mid-infrared range has been analyzed within the context of the quasiparticle peak approach in metallic systems.  It primarily arises from transitions between the coherent quasiparticle peak at the Fermi energy and the Hubbard bands (discussed later) \cite{Istvan.2004}.
Furthermore, we note that the dc conductivity as a function of temperature, as extracted from the optical conductivity data, fully agrees with the electric transport results presented earlier [see Fig.\ \ref{fig.parameters1}(d)]. Interestingly, the Kondo effect, which is revealed in the temperature-dependent $\sigma_{dc}$ data as a broad maximum at around 30~K, leads to a decrease in the Drude spectral weight $SW_{D}$ and an increase in the carrier scattering rate below $\sim$30~K. These changes might be related to the Kondo effect or anomalous Hall effect \cite{Nakatsuji.2006, harish.2021, Machida.2007}.

\begin{figure}[t]
	\includegraphics[width=1\linewidth]{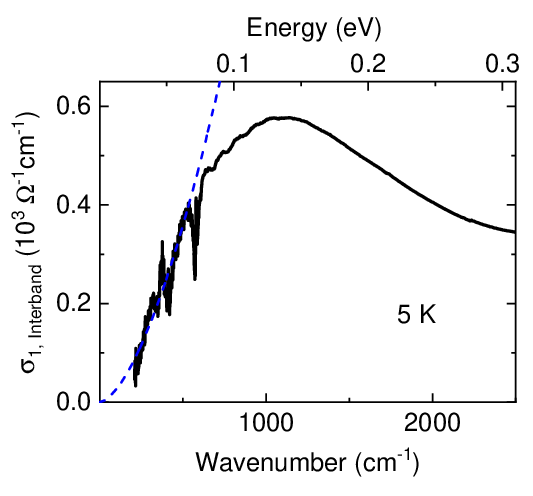}
	\caption{Interband optical conductivity $\sigma_{1,interband}$ at 5~K obtained by subtracting the Drude term (D) and the phonon mode contributions from the total $\sigma_{1}$ spectrum. The dashed line indicates a fit with a power law behavior as described in the text.
}
	\label{fig.MIRband}
\end{figure}

\section{Discussion}

According to our prior discussion, Pr$_2$Ir$_2$O$_7$ is an interesting material in the context of exotic topological states in pyrochlore iridates \cite{Pesin.2010,William.2014,Wan.2011,Moon.2013,Matsuhira.2011,Nakatsuji.2006}.
As illustrated in Figs.\ \ref{fig.reflectivity1} and \ref{fig.Fitting1}, the interband optical conductivity within the low-energy regime (mid-IR range) exhibits a pronounced temperature dependence throughout the entire temperature range.
The significance of novel topological phases such as 3D Weyl/Dirac materials and quadratic band touching is important in the current context \cite{Goswami.2017, Lee.2013, Kondo.2015, Ueda.2016, Sushkov.2015}. Generally, the interband optical conductivity in $d$ dimension obeys the universal formula $\sigma_{1} \propto \omega^{(d-2)/z}$, where $z$ represents the exponent in the band dispersion relation $E(k)$ $\propto$ $|k|^{z}$ \cite{Adam.2013}. In three-dimensional (3D, $d$ = 3) materials characterized by parabolic band dispersion ($z$ = 2), $\sigma_{1}$
follows a square root frequency dependence according to $\sigma_{1} \propto \omega^{0.5}$ \cite{Schilling.2018}.
Furthermore, the observed linear-in-frequency behavior of the interband optical conductivity in a 3D system serves as a distinct indication of linear band dispersion ($z$ = 1), a characteristic feature of WSM/Dirac materials \cite{Pavan.2012}.
A recent study of Nd$_2$Ir$_2$O$_7$ shows that the $\sigma_{1}(\omega)$ exhibits a linear-in-frequency ($\omega$) dependence in the low-energy region \cite{Wang.2020}.
Additionally, there is a notable transfer of spectral weight from the Drude to the interband component of $\sigma_{1}(\omega)$ as temperature decreases and the Drude spectral weight exhibits a $T^2$ dependence \cite{Wang.2020,Corasaniti.2021}. Furthermore, the application of an external magnetic field can modulate the valley population in the Weyl phase through $f-d$ coupling \cite{Kapon.2022}.

First, we note that we do not observe a $T^2$ dependence in the Drude spectral weight $SW_{D}$ [see Fig.\ \ref{fig.parameters1}(b)].
To further check the above point, we subtract the Drude term ($D$) and the phonon modes contributions from the total $\sigma_{1}(\omega)$ spectrum of 5 K data. The so-obtained interband conductivity $\sigma_{1,Interband}(\omega)$ is depicted in Fig.\ \ref{fig.MIRband}.
The spectral shape of the present {$\sigma_{1,Interband}(\omega)$} data does neither conform to the characteristic behavior of 3D quadratic band touching \cite{Wang.2020,Cheng.2017}, $\sigma_{1,Interband}(\omega)$ $\propto$ $\omega^{0.5}$, nor to the expected frequency dependence of 3D Weyl/Dirac cones \cite{Pavan.2012,Adam.2013,Wang.2020,Ueda.2016}, $\sigma_{1,Interband}(\omega)$ $\propto$ $\omega$.
We also tried the power law behavior $\sigma_{1,Interband}(\omega)$ $\propto$ $\omega^{n}$ (where, $n$ is the exponent) for the lowest measured 5 K $\sigma_{1,Interband}(\omega)$ spectra, as depicted in Fig.\ \ref{fig.MIRband}. The fit yields an exponent $n$ = 1.62(1), which closely aligns with the observed value for Nd$_2$Ir$_2$O$_7$ where it signals the emergence of the correlation-gapped WSM phase \cite{Wang.2020}. Intriguingly, in contrast to Nd$_2$Ir$_2$O$_7$ which undergoes a metal-to-insulator transition (MIT) and magnetic ordering at $T_N \sim 33$ K \cite{Matsuhira.2011}, Pr$_2$Ir$_2$O$_7$ maintains a PM metallic nature \cite{Nakatsuji.2006,harish.2021}. The observed  exponent $n$ suggests that Pr$_2$Ir$_2$O$_7$ is in proximity to the WSM phase. Pr$_2$Ir$_2$O$_7$ is conjectured to reside near an AFM QCP \cite{Savary.2014,Tokiwa.2014}. This QCP has been extensively explored as a crucial transition boundary between an AFM WSM and a phase characterized by a nodal non-Fermi liquid \cite{Savary.2014}. Additionally, Ueda \textit{et. al} found the signatures of topological transitions among the WSM states for Pr$_2$Ir$_2$O$_7$ under the influence of a magnetic field, arising notably from the $f-d$ coupling \cite{KUeda.2022}. Hence, the observed value of $n$ for Pr$_2$Ir$_2$O$_7$ implies the proximity of the WSM phase. This observation underscores the need for more comprehensive investigations to potentially transition the current material towards the WSM side through dedicated perturbation of the Pr-site with a smaller A-ionic size or by applying external pressure.

Next, we discuss the relevance of electronic correlations for the interpretation of the optical conductivity of Pr$_2$Ir$_2$O$_7$.
In many correlated materials, spectral weight due to intraband excitations, primarily located in the far-IR range, is significantly suppressed and transferred to higher frequencies in the mid- and near-IR \cite{Lee.2013}. The existence of the $M$-band in the optical conductivity of Pr$_2$Ir$_2$O$_7$ may thus be attributed to electron correlation effects.
The strength of electronic correlations in a material is calculated by the ratio of the kinetic energies K$_{opt}$/K$_{band}$, where K$_{opt}$ is the optical kinetic energy and K$_{band}$ the band kinetic energy \cite{Basov.2011}. According to Ref.\ \cite{Degiorgi.2011}
the ratio K$_{opt}$/K$_{band}$ can be obtained via the ratio of spectral weights $SW_{intra}$/$SW_{tot}$ from the experimental optical conductivity data.
The total spectral weight $SW_{tot}$ was computed by integrating $\sigma_{1}(\omega$) at 5~K up to 2700 cm$^{-1}$ (0.334 eV), covering both Drude term and MIR-band, and subtraction of the phonon mode spectral weight.
The intraband contribution $SW_{intra}$ was then determined by subtracting the fitted $M$-band contribution from $SW_{tot}$.
The ratio of $SW_{intra}$/$SW_{tot}$ for Pr$_2$Ir$_2$O$_7$ at 5~K amounts to approximately 0.26, representing a close low value compared to the value of 0.5 observed for its counterpart Bi$_2$Ir$_2$O$_7$ \cite{Lee.2013} and falling within the typical range for many correlated metals \cite{Lee.2013,Basov.2011}.

In contrast to the well-understood $A$, $B$, $B^{'}$, and $C$ excitations, the enigmatic presence of the $M$-band in the mid-IR region for the metallic nature of Pr$_2$Ir$_2$O$_7$ poses an intriguing puzzle. We described the $M$-band with a Lorentz oscillator positioned at $\sim$ 0.121 eV and $\sim$ 0.128 eV for the 300 K and 5 K, respectively, aligning closely with reported values for R-pyrochlore iridates \cite{Ueda.2016}.
Previously, the $M$-band was also observed in the other metallic pyrochlores Bi$_2$Ir$_2$O$_7$ and Pb$_2$Ir$_2$O$_7$, located at around $\sim$ 0.2 eV and $\sim$ 0.4 eV, respectively \cite{Lee.2013,Hirata.2013}. Notably, these energy values exceed the MIR-band position as observed in Pr$_2$Ir$_2$O$_7$. The origin of this mid-infrared peak in Bi$_2$Ir$_2$O$_7$ was suggested to be related to correlation-induced interband transitions.
Within the Ruddlesden-Popper (RP) series, Sr$_2$IrO$_4$ serves as a Mott insulator (MI) and displays a double-peak structure in the $\sigma_{1}(\omega$) spectrum around 0.5 and 1 eV \cite{Kim.2008, Moon.2008}. These peaks correspond to on-site and inter-site transitions between $J_{eff}$ = 1/2 and $J_{eff}$ = 3/2 states. As the system undergoes a transition from the insulating state (Sr$_2$IrO$_4$) to nearly insulating (Sr$_3$Ir$_2$O$_7$), and finally to metallic state (SrIrO$_3$) within the RP series, the $J_{eff}$ bands shift to lower energies \cite{Moon.2008}. Notably, $J_{eff}$ = 1/2 is observed around 0.2 eV for SrIrO$_3$ \cite{Moon.2008}, still higher in energy than the $M$-band in our $\sigma_{1}(\omega$) spectra.
This $J_{eff}$ shift is also observed in pyrochlores transitioning from insulating (Y$_2$Ir$_2$O$_7$) to MIT side (Sm$_2$Ir$_2$O$_7$) \cite{Ueda.2016}. In MIT-based Nd$_2$Ir$_2$O$_7$, an absorption band is observed, exhibiting a notable shift from approximately 1 eV to around 0.1 eV with lowering the temperature from 290 K to 5 K. Despite this shift, a charge gap of 45 meV remains evident at 5 K and goes to the zero-gap state by Rh doping ($x$) \cite{Ueda.2012}.
Although the expected magnitude of $U$ for 5$d$ Ir oxides is below 1 eV \cite{Moon.2008,Kim.2008,Hermann.2017,Kumar2022}, it's noteworthy that the bandwidth of the $J_{eff}$ = 1/2 band, shaped by spin-orbit coupling, may also exhibit narrow characteristics.

Moreover, in the context of the bandwidth during the transition from an MI to a metallic state, it has been observed that a quasi-particle (QP) peak tends to appear near the Fermi level at low temperatures $E_{F}$ \cite{Istvan.2004,Lee.2005,Lee.2001}, situated between the lower and upper Hubbard bands, as illustrated in Fig.\ \ref{fig.Scheme}. As the system undergoes a transition towards a more metallic state, the QP peak is expected to intensify with reductions in the Hubbard bands.
Accordingly, one expects several contributions in the optical conductivity spectrum, namely (i) a coherent peak centered at zero energy, (ii) an absorption band due to electronic excitations between the Hubbard bands and the QP peak, and (iii) robust $p-d$ transitions situated at higher energies \cite{Lee.2005}. Furthermore, within the Hubbard model it is expected that the spectral density of the QP band increases with lowering the temperature \cite{Merino.2000}.

In our data we observe the expected spectral features along with a moderate increase in both mode strength $S_{M}$ of the $M$-band and $SW_{D}$ of the Drude term as the temperature decreases. This phenomenon could potentially be attributed to the QP peak feature in the present system, where the transitions take place from the lower Hubbard band to the QP peak or from the QP peak to the upper Hubbard band, as indicated by the green arrow and labelled by $M$ in Fig.\ \ref{fig.Scheme}. We, however, note that the spectral weight transfer from high to low energies [see Fig.\ \ref{fig.reflectivity1}(b)] cannot be explained within this scenario.

Finally, we note that a midinfrared absorption band is observed in the optical conductivity spectrum of many transition-metal oxides and has been attributed to the excitations of polarons formed due to strong electron-phonon coupling \cite{Lupi.1999,Bi.1993,Kuntscher.2003,Ebad-Allah.2012,Devreese.2010,Reticcioli.2018}.
Also for the closely related metallic pyrochlore iridate Bi$_2$Ir$_2$O$_7$ the possible relevance of electron-phonon coupling effects has been discussed \cite{Lee.2013}.
Further theoretical investigations are needed to unravel the impact of electronic correlations and electron-phonon coupling on the low-frequency optical response in metallic and semi-metallic iridates.

\section{Conclusion}
In conclusion, our investigation of the pyrochlore iridate Pr$_2$Ir$_2$O$_7$ has provided insights into its magnetic, lattice dynamical, and electronic properties. Raman spectroscopy confirmed the cubic \textit{Fd$\bar{3}$m} crystal symmetry, while dc magnetic susceptibility exhibits the paramagnetic nature of Pr$_2$Ir$_2$O$_7$. Resistivity measurements show its metallic behaviour, further supported by optical conductivity analysis. Notably, the optical conductivity spectrum featured a mid-infrared absorption band, intensifying at lower temperatures due to spectral weight redistribution, highlighting the role of correlation effects. Furthermore, our optical response analysis suggests that Pr$_2$Ir$_2$O$_7$ lies in proximity to a Weyl semimetal phase.

\begin{acknowledgments}
H.K. acknowledges support provided by the Deutsche Forschungsgemeinschaft, Germany, under Grant No. KU 4080/2-1 (495076551). P.T. was supported by the Alexander von Humboldt foundation. This work was funded by the Deutsche Forschungsgemeinschaft (DFG, German Research Foundation) –- TRR 360 –- 492547816.
\end{acknowledgments}

\end{document}